\documentclass[aps,prl,twocolumn,showpacs,superscriptaddress,groupedaddress]{revtex4} 
\usepackage[utf8]{inputenc}
\usepackage{hyperref}
\hypersetup{
    colorlinks=true,
    linkcolor=blue,
    filecolor=magenta,      
    urlcolor=cyan,
    pdfpagemode=FullScreen,
}
\usepackage{verbatim}
\usepackage{upgreek}
\usepackage{amsmath,amssymb,amsopn}

\usepackage{braket}
\usepackage{graphicx,animate}
\usepackage[dvipsnames]{xcolor}
\usepackage{soul,xcolor}
\usepackage{xcolor, ulem, color, framed}

\begin{document}

\title{Heavy quark drag and diffusion coefficients in the pre-hydrodynamic QCD plasma}

\author{Xiaojian Du} 
\email{xiaojian.du@usc.es}
\affiliation{Instituto Galego de Física de Altas Enerxías (IGFAE), Universidade de Santiago de Compostela, E-15782 Galicia, Spain}
\date{\today}

\begin{abstract}
Kinetic and chemical equilibrations play important roles in the formation of the quark-gluon plasma (QGP) in relativistic heavy-ion collisions (HICs). 
These processes further influence the production of hard and electromagnetic probes in HICs, in particular, the thermalization of heavy quarks, which are produced at an extremely early time before the formation of the QGP.
We calculate the drag and diffusion coefficients of heavy quarks in the pre-hydrodynamic quantum chromodynamic (QCD) plasma with the state-of-the-art QCD effective kinetic theory (EKT) solver.
We present the time, momentum, and angular dependencies of these coefficients for gluon and quark contributions separately, showing the effects of isotropization and chemical equilibration from the QCD plasma.
We also provide a simple formula to estimate the heavy quark drag and diffusion coefficients, as well as its energy loss, within the pre-hydrodynamic plasma at different coupling strengths based on the attractor theory.
We then discuss the validity of these estimations with leading-order calculations and leading-logarithmic rescaling factors.
\end{abstract}

\maketitle

\textbf{Introduction}

Thermalization is omnipresent and there are two main categories of thermalization systems that are being widely studied due to the simplification of degrees of freedom at certain scales.
One is the thermalization of many-body closed quantum systems, where the microscopic dynamics of single particles can be coarse-grained and the emergent behavior is more interesting.
The other one is the thermalization of open quantum systems, where the objects have distinguished degrees of freedom or scales from the background medium environment, and tracing out the environment leaves a simple dynamical description of the system in the medium.
Relativistic heavy-ion collisions (HICs) are such experiments that both categories are present for us to understand the fundamental strong interaction.

The quark-gluon plasma (QGP) containing free quark and gluon degrees of freedom, as a many-body system, can only be produced in the early universe or HICs nowadays.
The main period of the QGP evolution in HICs is successfully described by a near-equilibrium macroscopic theory, the relativistic hydrodynamics~\cite{Muller:1967zza,Israel:1979wp,Baier:2007ix,Luzum:2008cw,Schenke:2010rr,Song:2010mg,Pang:2012he,Du:2019obx}, in terms of the energy-momentum tensor.
A more involved tool that can describe the non-equilibrium and pre-hydrodynamic QGP is the effective kinetic theory (EKT)~\cite{Arnold:2002zm}, in terms of particle distributions.
This theory is initially implemented as Yang-Mills kinetics~\cite{Kurkela:2014tea,Kurkela:2015qoa} including gluon, and developed into quantum chromodynamic (QCD) kinetics~\cite{Kurkela:2018oqw,Du:2020dvp} including both gluon and quarks.
Various models as similar approaches or as simplified versions of EKT exist~\cite{Romatschke:2003ms,Xu:2004mz,Martinez:2010sc,Blaizot:2013lga,Kamata:2020mka,Behtash:2020vqk,Ambrus:2022qya,BarreraCabodevila:2022jhi,Brewer:2022ifw}.
Although the pre-hydrodynamic QGP in HICs is complicated as is anisotropic and chemically out-of-equilibrium, there are still some universal descriptions of the pre-hydrodynamic QGP based on simple conservation laws, independent of the microscopic physics, such as the attractor theory~\cite{Heller:2015dha,Romatschke:2017vte,Strickland:2018ayk,Kurkela:2019set,Giacalone:2019ldn,Du:2020zqg,Heller:2020anv,Almaalol:2020rnu,Chattopadhyay:2021ive,Du:2022bel}.

On the other hand, heavy quarks have distinguished mass scales from light partons in the QGP, and are produced as open quantum systems due to their large mass thresholds.
Heavy quark thermalization is contributed by energy loss and diffusion.
Most of the heavy quarks are produced within the momentum range $p\lesssim m_{\rm HQ}$, where the radiative energy loss can be neglected and the collisional energy loss dominates~\cite{Moore:2004tg}. 
They are produced at a time scale $\tau\simeq \frac{1}{m_{\rm HQ}}$ before the hydrodynamization of the QGP at $\tau_h\simeq \frac{4\pi\eta}{Ts}$, and relax at a much later time $\tau_R\simeq\frac{m_{\rm HQ}}{T}\tau_{h}$.
Thus, most of the heavy quark thermalization simulations~\cite{Svetitsky:1987gq,GolamMustafa:1997id,vanHees:2004gq,Riek:2010fk,Cao:2011et,He:2011qa,Lang:2012nqy,Das:2013kea,Aichelin:2013mra,Beraudo:2014boa,Song:2015sfa,Cao:2016gvr,Kang:2016ofv,Ke:2018tsh,He:2022ywp,Du:2022uvj} are focusing on the hydrodynamic stage when the QGP is nearly thermalized.
There are some efforts in addressing the heavy quark thermalization in the pre-hydrodynamic glasma or QGP~\cite{Romatschke:2004au,Adare:2013wca,Sun:2019fud} taking care of the anisotropy or chemical effects.
The EKT allows us to calculate the heavy quark diffusion~\cite{Boguslavski:2023fdm}  as well as jet momentum broadening~\cite{Boguslavski:2023alu} dynamically during the pre-hydrodynamic stage from the first principle.
Furthermore, the recent developments of the EKT to a full QCD level will complete this picture.

In this article, we will show the first-principle calculations of heavy quark drag and diffusion coefficients in the QCD plasma, from the state-of-the-art QCD effective kinetic theory (QCD EKT) solver~\cite{Du:2020dvp} including both gluon and quark dynamics at fixed coupling strengths.
Since the pre-hydrodynamic QCD plasma undergoes a rapid drop in temperature and a transition from weakly coupled to strongly coupled, the interpolation from the weakly coupled plasma to the strongly coupled plasma as a realistic medium profile would be favored. Although the EKT breaks down at strong coupling, the universal attractor can perform such an interpolation regardless of the coupling strength.
It also avoids running complex EKT simulations at various coupling strengths by providing the rescaling of key medium profiles like the time $\tau$ and the temperature $T$ which are essential ingredients for heavy quark thermalization.
Augmented with a proper rescaling factor for the heavy quark diffusion, one would arrive at a realistic heavy quark energy loss estimation in the pre-hydrodynamic stage.
In this paper, we perform the first step towards this goal by justifying the validity of the rescaling at weak coupling and demonstrating the breakdown of the rescaling at strong coupling with leading-order (LO) perturbative QCD (pQCD) calculations and leading-logarithmic (LL) rescaling factors.

\textbf{Pre-hydrodynamic QCD plasma \& attractor}

The QCD plasma out-of-equilibrium before the formation of the hydrodynamic state can be described by the QCD EKT, with a Bjorken expansion at the early stage of HICs. 
Within the EKT, the evolution of gluon and light quark/anti-quarks as constitutes of the QCD plasma is formulated as a set of coupled Boltzmann equations~\cite{Mueller:1999pi} with $a=g,q,\bar{q}$ and flavor number $N_f=3$
\begin{eqnarray}
\frac{\partial f_{a}(\vec{p},\tau)}{\partial\tau}-\frac{p_{\|}\partial f_{a}(\vec{p},\tau)}{\tau\partial p_{\|}}=C_{a}^{1\leftrightarrow2,2\leftrightarrow2}[f](\vec{p},\tau).
\end{eqnarray}
The expansion term with prefactor $1/\tau$ anisotropizes the plasma 
at an early time, while the collision terms $C_{a}[f]$ take over the evolution at a later time and drive the plasma to reach hydrodynamic equilibrium, in both kinetic and chemical sense.
Details of the QCD EKT and its numerical implementations can be found in our previous paper~\cite{Du:2020dvp}.

\begin{figure}[t!]
	\centering
	\includegraphics[width=0.45\textwidth]
	{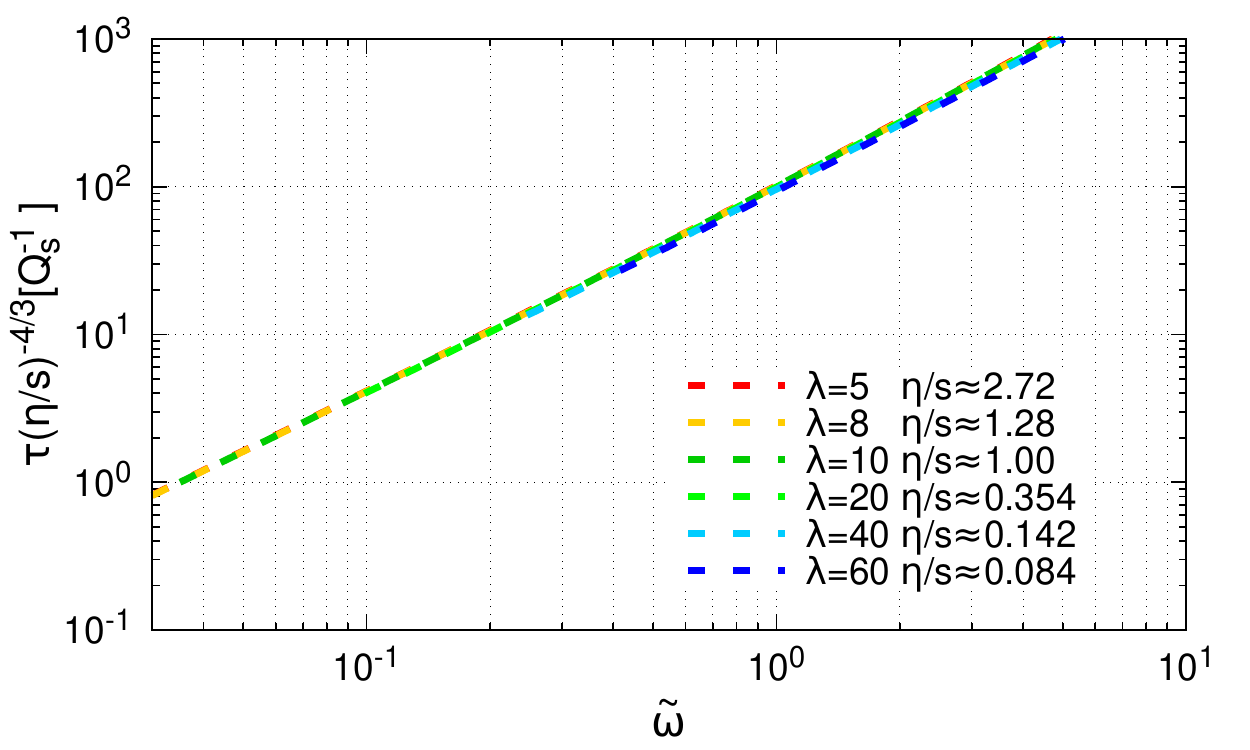} 
	\includegraphics[width=0.45\textwidth]
	{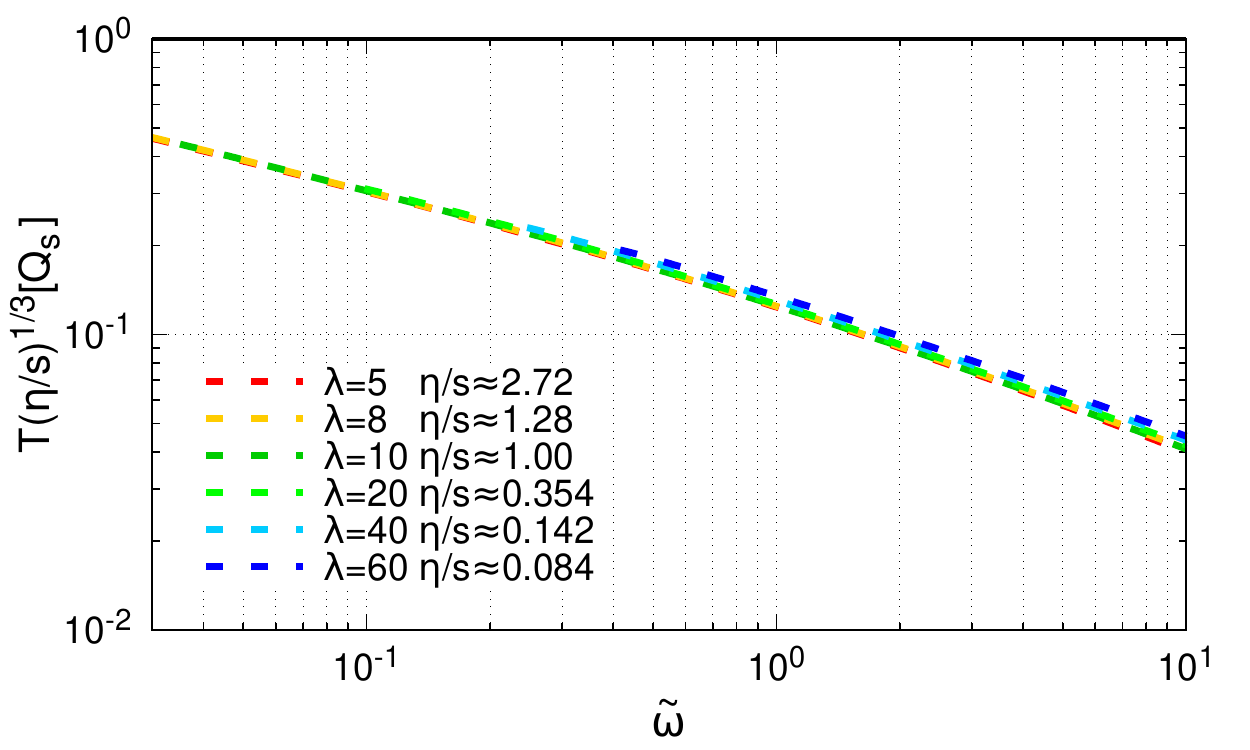} 
	\caption{Comparison of the rescaled time $\tau(\eta/s)^{-4/3}$ and temperature $T(\eta/s)^{1/3}$ according to Eq.~(\ref{eq:rescaling-tT}) at various 'tHooft couplings $\lambda=5,8,10,20,40,60$ from the QCD EKT simulations.}
	\label{fig:Plasma_Rescaling}
\end{figure}
The early-time expansion and the later-on hydrodynamization are independent of the microscopic interactions in the kinetic theory, resulting in a universal attractor solution. 
This solution connects the energy density of the plasma at any time $e(\tau)$ to its initial value $e_0$ in a simple and universal way~\cite{Giacalone:2019ldn,Du:2020zqg}
\begin{eqnarray}
\label{eq-attractor}
\tau^{\frac{4}{3}}e(\tilde{\omega})=\left(4\pi\frac{\eta}{s}\right)^{\frac{4}{9}}\left(\frac{\pi^2\nu_{\rm eff}}{30}\right)^{\frac{1}{9}}\left(\tau_0e_0\right)^{\frac{8}{9}}C_{\infty}\mathcal{E}(\tilde{\omega}).
\end{eqnarray}
The function $\mathcal{E}(\tilde{\omega})$ is called the energy attractor
in terms of the universal and dimensionless time scale $\tilde{\omega}=\frac{\tau Ts}{4\pi\eta}$. 
The effective temperature can be evaluated by Landau matching $T=\left(\frac{30e(\tau)}{\pi^2\nu_{\rm eff}}\right)^{\frac{1}{4}}$.
The degeneracy factor $\nu_{\rm eff}=\nu_g+\frac{7}{4}\nu_qN_f=47.5$ and $C_{\infty}=0.87$ are both constants. 
The shear viscosity over entropy density ratio $\eta/s$ directly reflects how strong the interaction is and how quick the equilibration can be. 
Indeed, at any $\tilde{\omega}$, there is a universal energy attractor $\mathcal{E}(\tilde{\omega})$ that characterizes the degree of thermalization, valued from $\mathcal{E}(\tilde{\omega}\to 0)=0$ to $\mathcal{E}(\tilde{\omega}\to\infty)=1$. 
One can evaluate the corresponding time $\tau$ and temperature $T$ at any universal time $\tilde{\omega}$ as
\begin{eqnarray}
\label{eq-time}
\nonumber
\tau=\left(4\pi\frac{\eta}{s}\right)_{\rm prehydro}^{\frac{4}{3}}\left(\frac{\pi^2\nu_{\rm eff}}{30}\right)^{\frac{1}{3}}\left(\tau_0e_0\right)^{-\frac{1}{3}}C_{\infty}^{-\frac{3}{8}}\mathcal{E}^{-\frac{3}{8}}(\tilde{\omega})\tilde{\omega}^{\frac{3}{2}},\\
T=\left(4\pi\frac{\eta}{s}\right)_{\rm prehydro}^{-\frac{1}{3}}\left(\frac{\pi^2\nu_{\rm eff}}{30}\right)^{-\frac{1}{3}}\left(\tau_0e_0\right)^{\frac{1}{3}}C_{\infty}^{\frac{3}{8}}\mathcal{E}^{\frac{3}{8}}(\tilde{\omega})\tilde{\omega}^{-\frac{1}{2}}.
\end{eqnarray}
This means that a more strongly coupled plasma with a smaller $\eta/s$ requires a shorter time to reach a certain degree of thermalization, while a more weakly coupled plasma with a larger $\eta/s$ requires a longer time. With a fixed initial energy density, the shorter thermalization time in strongly coupled plasma also results in higher initial temperatures for the following hydrodynamics.
As a consequence, the universality of the attractor gives the approximate relation in the pre-hydrodynamic stage for varying coupling strengths
\begin{eqnarray}
\label{eq:rescaling-tT}
\nonumber
\tau_{\rm prehydro}\left(\frac{\eta}{s}\right)^{-\frac{4}{3}}_{\rm prehydro}
&\simeq& {\rm coupling~independent},\\
T_{\rm prehydro}\left(\frac{\eta}{s}\right)^{\frac{1}{3}}_{\rm prehydro}
&\simeq& {\rm coupling~independent}.
\end{eqnarray}
The corresponding $\eta/s$ can be extracted from fitting to hydrodynamic constitutive relation as it was done in~\cite{Du:2020dvp}.
Fig.~\ref{fig:Plasma_Rescaling} shows comparisons of the rescaled time $\tau(\eta/s)^{-4/3}$ and temperature $T(\eta/s)^{1/3}$ at various 'tHooft coupling $\lambda=g^2N_c$ from the QCD EKT simulations. 
One observes the rescaling relation of Eq.~(\ref{eq:rescaling-tT}), even up to large coupling $\lambda=60$ where the $\eta/s\simeq 0.084$ is already close to the lower bound from holography $\eta/s=1/4\pi$~\cite{Kovtun:2004de}, although slight deviation is observed since the kinetic theory should break down at such a large coupling.

There is only one scale in a conformal theory and one can fix it by matching experimental data at the end of the QGP evolution.
In equilibrium, $\mathcal{E}(\tilde{\omega}\gg 1)=1$ and $sT=e+p$ at zero net-baryon density. The equation of state $e=3p$ always holds in a conformal theory. 
One has the entropy density in equilibrium
\begin{eqnarray}
\label{eq-entropy}
(\tau s)_{\rm eq}=\frac{4}{3}\frac{(\tau^{\frac{4}{3}}e)_{\rm eq}}{(\tau^{\frac{1}{3}}T)_{\rm eq}}=\frac{4}{3}(\tau^{\frac{4}{3}}e)_{\rm eq}^{\frac{3}{4}}\left(\frac{\pi^2\nu_{\rm eff}}{30}\right)^{\frac{1}{4}},
\end{eqnarray}
which is related to the charged particle multiplicity via $dN_{\rm ch}/d\eta=\frac{N_{\rm ch}}{S}(\tau s)_{\rm eq}S_{\perp}$ with $S/N_{\rm ch}=8.36$~\cite{Hanus:2019fnc} and $S_{\perp}$ the transverse area of the collision. 
As a consequence, one can constrain the initial condition $\tau_0e_0$ with the following relation~\cite{Giacalone:2019ldn}
\begin{eqnarray}
\label{eq:matching}
\frac{dN_{\rm ch}}{d\eta}=\frac{4}{3}\frac{N_{\rm ch}}{S}\left(4\pi\frac{\eta}{s}\right)_{\rm average}^{\frac{1}{3}}\left(\frac{\pi^2\nu_{\rm eff}}{30}\right)^{\frac{1}{3}}\left(\tau_0e_0\right)^{\frac{2}{3}}C_{\infty}^{\frac{3}{4}}S_{\perp}.
\end{eqnarray}
The average $\eta/s$ represents a measure of the thermalization speed for both the pre-hydrodynamic stage and the following hydrodynamic stage until freeze-out. It is, however, dominated by the hydrodynamic stage due to a much shorter lifetime of the pre-hydrodynamic period compared to the hydrodynamic period. A strongly coupled fluid dynamic features a small $\eta/s$ close to a holographic bound $\eta/s=1/4\pi$~\cite{Kovtun:2004de}.
Matching the LHC 5.02\,TeV Pb+Pb collision data~\cite{ALICE:2015juo}, $dN_{\rm ch}/d\eta=1942$ and $S_{\perp}=138$fm$^2$ to Eq.~(\ref{eq:matching}), one arrives at $\tau_0e_0=1.961$GeV$^3$ in this specific collision system. One has further the rescaling formula to evaluate the initial energy density in other circumstances
\begin{eqnarray}
\tau_0e_0=1.961{\rm GeV}^3\left(\frac{\frac{dN_{\rm ch}}{d\eta}}{1942}\right)^{\frac{3}{2}}\left(4\pi\frac{\eta}{s}\right)_{\rm average}^{-\frac{1}{2}}\left(\frac{S_{\perp}}{138{\rm fm}^2}\right)^{-\frac{3}{2}}.
\end{eqnarray}

The initial energy is mainly deposited by an over-occupied, anisotropic, and gluon-saturated state; and described by the CGC-inspired distribution~\cite{Lappi:2011ju,Kurkela:2015qoa}
\begin{eqnarray}
\nonumber
&&f_{g}(\vec{p},\tau_0)=\frac{10.5}{\lambda_0}\frac{1.8Q_s}{\sqrt{p_{\perp}^2+(\xi p_{\|})^2}}\exp \left[-\frac{2}{3}\frac{p_{\perp}^2+(\xi p_{\|})^2}{(1.8Q_s)^2}\right],\\
&&f_{q}(\vec{p},\tau_0)=f_{\bar{q}}(\vec{p},\tau_0)=0,
\end{eqnarray}
with the momentum decomposition in transverse and longitudinal directions $\vec{p}=(\vec{p}_{\perp},p_{\|})$.
The anisotropic parameter is typically chosen as $\xi=10$. 
A typical value for the 'tHooft coupling $\lambda_0=g_0^2N_c$ chosen in simulating weakly coupled gauge field in the CGC effective theory is $\lambda_0=10$, which is 
reasonable as well at the initial time for our QCD kinetic simulation when the system has a high temperature.
The general 'tHooft coupling $\lambda=g^2N_c$ entering into the QCD kinetic simulation is controlling the thermalization speed, which can be further reflected in the macroscopic coefficient $\eta/s$. 
We keep the weak coupling $\lambda=10$ for the QCD plasma as the default throughout the QCD kinetic simulation and perform rescaling to evaluate strongly coupled plasma, where the validity of both the kinetic theory and the perturbation theory breaks down.
General rescaling can be achieved due to the universality of the attractor solutions, from the basic principles of energy conservation and conformality, regardless of the coupling or modeling.
To extend the rescaling from the QCD plasma to the heavy quark, we will discuss the validity of rescaling for the transport coefficients in a standalone section later.
The initial time for the QCD kinetic evolution is approximately the inverse of the saturation scale for the gauge fields $\tau_0\simeq 1/Q_s$ before the formation of the quasi-particles in the kinetic theory. 
Without losing generality, by choosing $\tau_0=Q_s^{-1}$ we can calculate the initial energy density and one gets $\tau_0e_0=0.5858Q_s^3=1.961{\rm GeV}^3$. 
Now one can estimate that $Q_s=1.496{\rm GeV}$ and $\tau_0=Q_s^{-1}=0.134{\rm fm}$, smaller than the typical hydrodynamization time $\tau_{\rm h}\simeq 0.2-0.6{\rm fm}$ but at the same order of magnitude.

\begin{figure}[t!]
    \centering
    \includegraphics[width=0.45\textwidth]
    {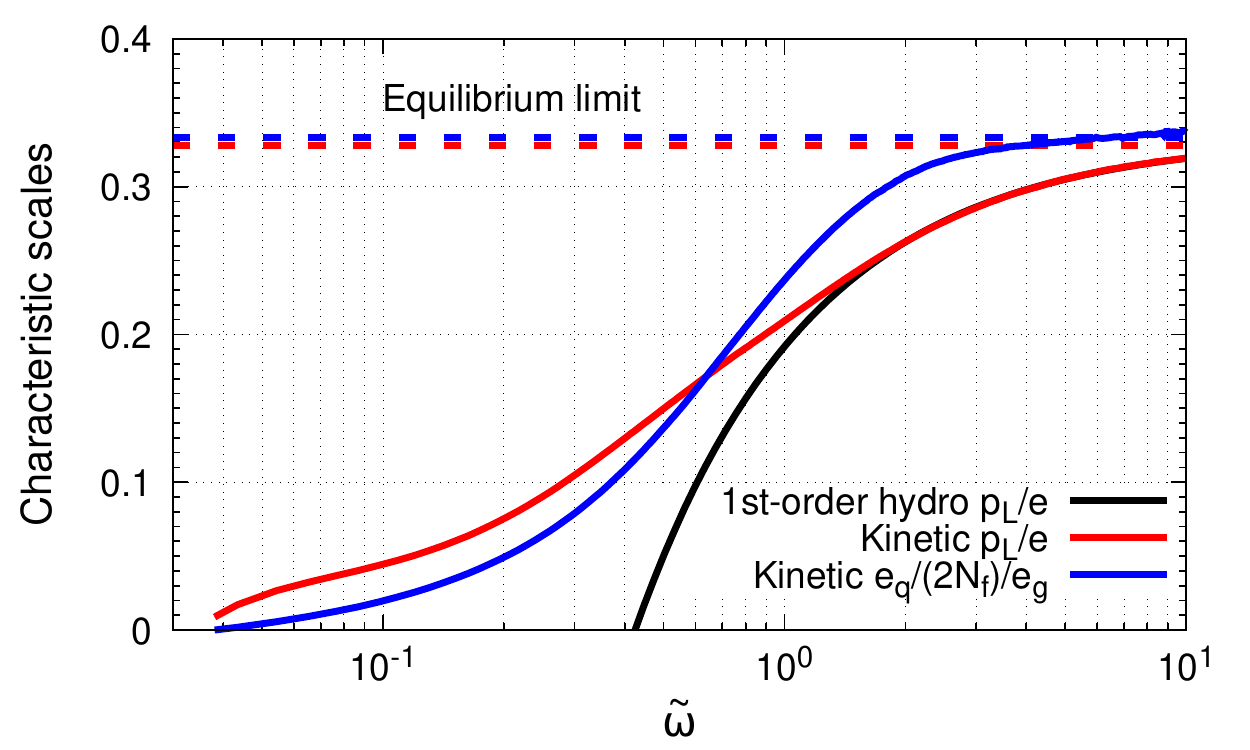} 
    \caption{Typical characteristic scales of isotropization $p_L/e$ (red) and chemical equilibration $e_q/e_g$ (blue) in terms of universal time $\tilde{\omega}$, compared to the hydrodynamic limit (black) and their equilibrium limits (dashed).  These curves and related discussions can also be found in our previous article on QGP thermalization~\cite{Du:2020dvp}.}
    \label{fig:Plasma_time}
\end{figure}

By solving the QCD EKT, one gets the time evolution of the distributions $f_{g,q,\bar{q}}(\vec{p},\tau)$. 
Certain quantities can characterize the equilibration of the QCD plasma, such as the energy-momentum tensor
\begin{eqnarray}
T^{\mu\nu}
=\int\frac{d^3p}{(2\pi)^3}\frac{p^{\mu}p^{\nu}}{p}
\left\{\nu_gf_{g}(\vec{p})+\nu_q N_f\left[f_{q}(\vec{p})+f_{\bar{q}}(\vec{p})\right]\right\}.
\end{eqnarray}
The longitudinal pressure over energy density ratio $p_L/e=T^{zz}/T^{\tau\tau}$ characterizes the isotropization of the plasma with an equilibrium limit $1/3$. 
The quark over gluon energy density ratio $e_q/e_g$ characterizes the chemical equilibration of the plasma with an equilibrium limit $(7\nu_{q}N_f)/(4\nu_g)$.
We show these characteristic scales in Fig.~\ref{fig:Plasma_time} in terms of the universal time scale $\tilde{\omega}$ for the QCD plasma at constant coupling $\lambda=10$.
The anisotropy of the plasma approaches the hydrodynamic limit at around $\tilde{\omega}\simeq 1-2$ while its equilibrium limit has to be reached after a much longer time.
The chemical equilibration roughly finishes later than $\tilde{\omega}\simeq 2-3$ where the quark over gluon density ratio tends to be a plateau.

These non-equilibrium parton distributions will deviate the heavy quark transport coefficients from thermal cases, from the initial time $\tau_0\simeq1/Q_s$ to the hydrodynamization time $\tau_h(\tilde{\omega\simeq 1-2})$, before the formation of the hydrodynamic plasma. This opens an opportunity to extend heavy quark simulations to the pre-hydrodynamic stage of HICs.

\textbf{Heavy quark thermalization}

The heavy quark thermalization with soft collisions from the background QCD plasma can be described by a stochastic differential equation (Langevin) in phase space $(\vec{x},\vec{p})$
\begin{eqnarray}
\nonumber
dx_{i}&=&\frac{p_i}{E(\vec{p})}d\tau\,,\\
dp_{i}&=&-A_i(\vec{p},\tau)d\tau+\sigma_{ij}(\vec{p},\tau)dW_j\,,
\label{eq-langevin}
\end{eqnarray}
with a Wiener process $dW_j\sim\mathcal{N}(0,d\tau)$ correlated as $\braket{dW_idW_j}=\delta_{ij}d\tau$.
Applying Ito's lemma to Eq.~(\ref{eq-langevin}) up to order $\mathcal{O}(d\tau)$, the Kolmogorov equation (Fokker-Planck) reads
\begin{eqnarray}
\label{eq-fokkerplank}
\frac{\partial f_Q(\vec{p},\tau)}{\partial\tau}
=\frac{\partial[A_i(\vec{p},\tau)f_Q(\vec{p},\tau)]}{\partial p_{i}}+\frac{\partial^2[B_{ij}(\vec{p},\tau)f_Q(\vec{p},\tau)]}{\partial p_{i}\partial p_{j}}\,.
\end{eqnarray}
There are two evolving chemical contributions for the drag coefficients $A_{i}(\vec{p},\tau)$ and diffusion coefficients $B_{ij}(\vec{p},\tau)=\frac{1}{2}\sigma_{ik}(\vec{p},\tau)\sigma_{jk}(\vec{p},\tau)$, from gluon collisions $gQ\rightarrow gQ$ and from quark/antiquark collisions $qQ\rightarrow qQ$ 
(including the factor $2N_f$ in the quark sector)
\begin{eqnarray}
\nonumber
A_{i}(\vec{p},\tau)&=&A_{g,i}(\vec{p},\tau)+A_{q,i}(\vec{p},\tau)\,,\\
B_{ij}(\vec{p},\tau)&=&B_{g,ij}(\vec{p},\tau)+B_{q,ij}(\vec{p},\tau)\,.
\end{eqnarray}

The drag and diffusion coefficients for heavy quark $Q$ collided by a parton $a=g,q,\bar{q}$ in the QCD plasma can be calculated as~\cite{Svetitsky:1987gq}
\begin{eqnarray}
\nonumber
&&A_{a,i}(\vec{p},\tau)=\frac{1}{2E(\vec{p})}\int d\Pi \overline{|M_{aQ\to aQ}|^2}\nu_a f_a(\vec{p}_a,\tau)\\
\nonumber
&&\times(1\pm f_a(\vec{p}_a',\tau))(1-f_Q(\vec{p}',\tau))(\vec{p}-\vec{p}')_i\,,\\
\nonumber
&&B_{a,ij}(\vec{p},\tau)=\frac{1}{4E(\vec{p})}\int d\Pi \overline{|M_{aQ\to aQ}|^2}\nu_a f_a(\vec{p}_a,\tau)\\
\nonumber
&&\times(1\pm f_a(\vec{p}_a',\tau))(1-f_Q(\vec{p}',\tau))(\vec{p}-\vec{p}')_i(\vec{p}-\vec{p}')_j\,,\\
\nonumber
{\rm with}&&d\Pi=\frac{d^3p_a}{(2\pi)^3 2E_{a}(\vec{p}_a')}\frac{d^3p_a'}{(2\pi)^3 2E_{a}'(\vec{p}_a')}\frac{d^3p_Q'}{(2\pi)^3 2E'(\vec{p}')}\\
&&\times (2\pi)^4\delta^{(4)}(P+P_a-P'-P_a')\,.
\end{eqnarray}
Due to the small occupation of heavy quarks $f_Q(\vec{p}',\tau)\ll 1$, the Fermi-blocking factor for the heavy quark can be neglected.
We calculate the amplitude squares $\overline{|M_{aQ\rightarrow aQ}|^2}$ of heavy quark scattering in LO pQCD, with a dynamical and isotropic screening mass fitting to the Hard Thermal Loop (HTL) calculation, the same as it is implemented in the QCD EKT simulation for QCD plasma~\cite{Arnold:2002zm}. For details, see our previous work~\cite{Du:2020dvp}.

To calculate the drag and diffusion coefficients, we assume a charm quark mass $m_{\rm HQ}=1.5$\,GeV, and the coupling in the collisional amplitude squares the same as the background QCD plasma $\alpha_s=\frac{g^2}{4\pi}=\frac{\lambda}{4\pi N_c}$ with $\lambda=10$ if not claimed otherwise.

\begin{figure}[t!]
	\centering
	\includegraphics[width=0.45\textwidth]
	{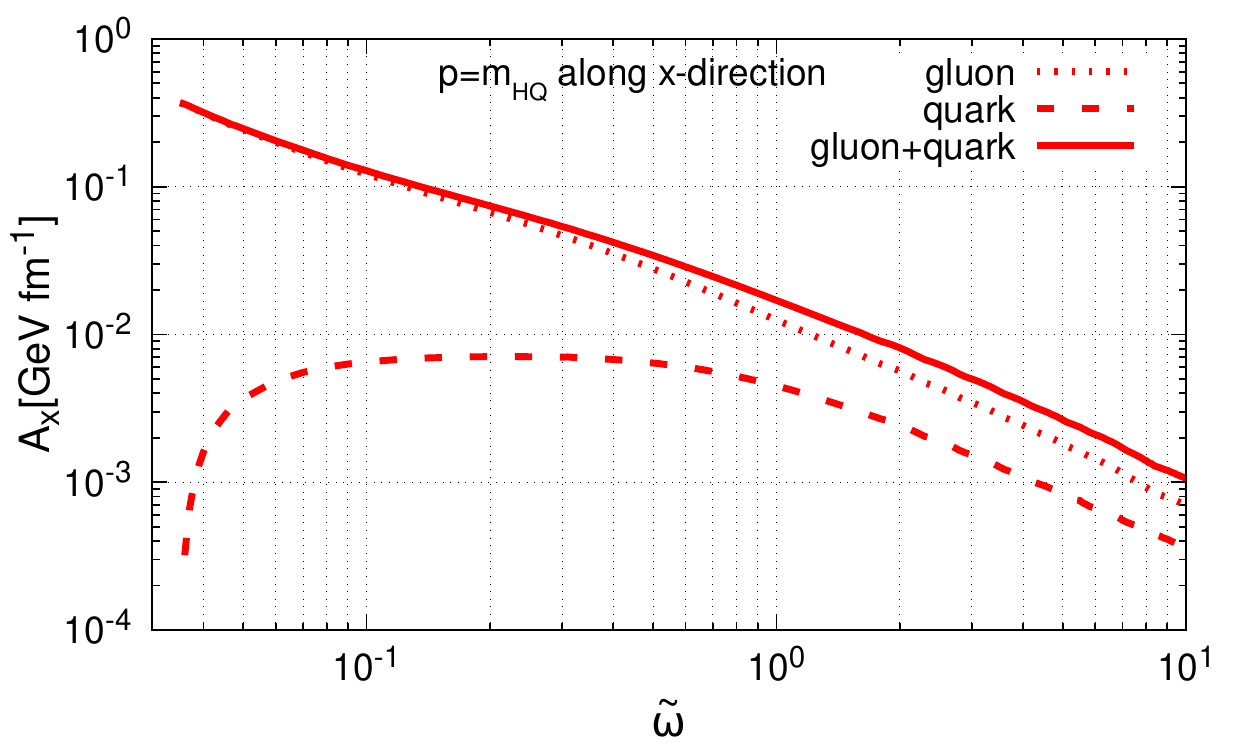} 
	\includegraphics[width=0.45\textwidth]
	{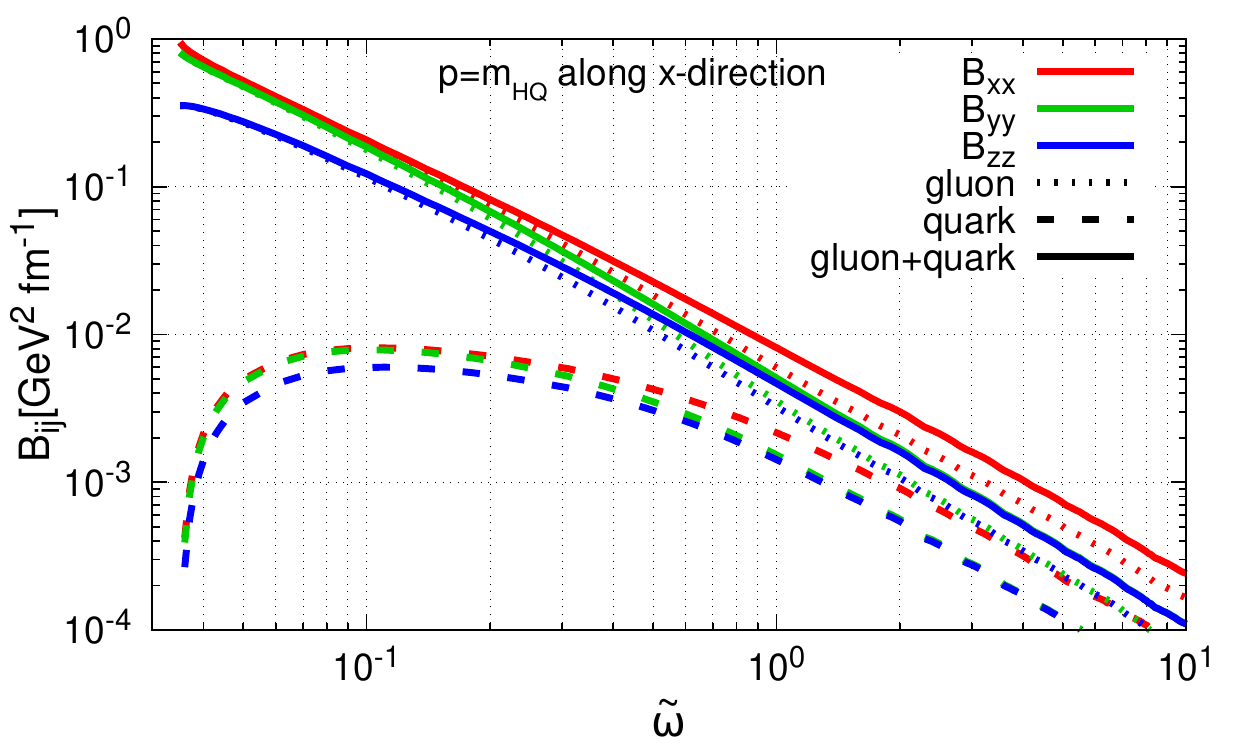} 
	\caption{Drag and diffusion coefficients $A_{x}(\tilde{\omega})$, $B_{xx}(\tilde{\omega})$ (red), $B_{yy}(\tilde{\omega})$ (green), $B_{zz}(\tilde{\omega})$ (blue) for gluon and quark (with factor 2$N_f$), as a function of universal time $\tilde{\omega}$, with heavy quark momentum $p=m_{\rm HQ}$ and mass $m_{\rm HQ}=1.5$\,GeV. The default coupling $\lambda=10$ is in use. Gluon and quarks are plotted as dotted and dashed curves respectively.}
	\label{fig:AB_time}
\end{figure}

\textbf{Heavy quark drag and diffusion coefficients}

Due to the rotation symmetry in the transverse plane, without loss of generality, we define the momentum direction of the heavy quark in the transverse plane as the x-axis.
That is $\vec{p}=(p,\cos(\theta),\phi)=(p,0,0)$ in cylindrical coordinate.
The symmetry of the integration makes the vector $\vec{A}$ simply along $\vec{p}$ in transverse plane, giving trivial values of $A_{y}$, $A_{z}$, as well as off-diagonal terms in the $B_{ij}$ matrix, even if the plasma is anisotropic.

\begin{figure}[t!]
	\centering
	\includegraphics[width=0.45\textwidth]
	{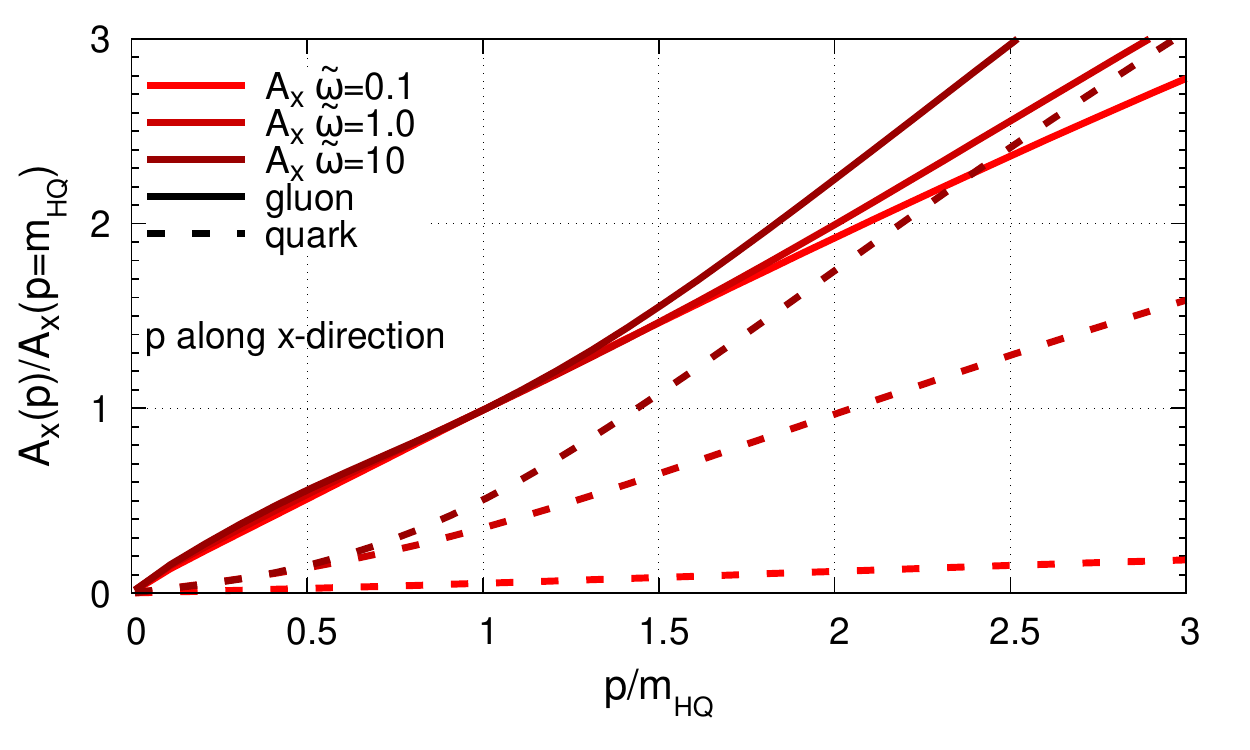} 
	\includegraphics[width=0.45\textwidth]
	{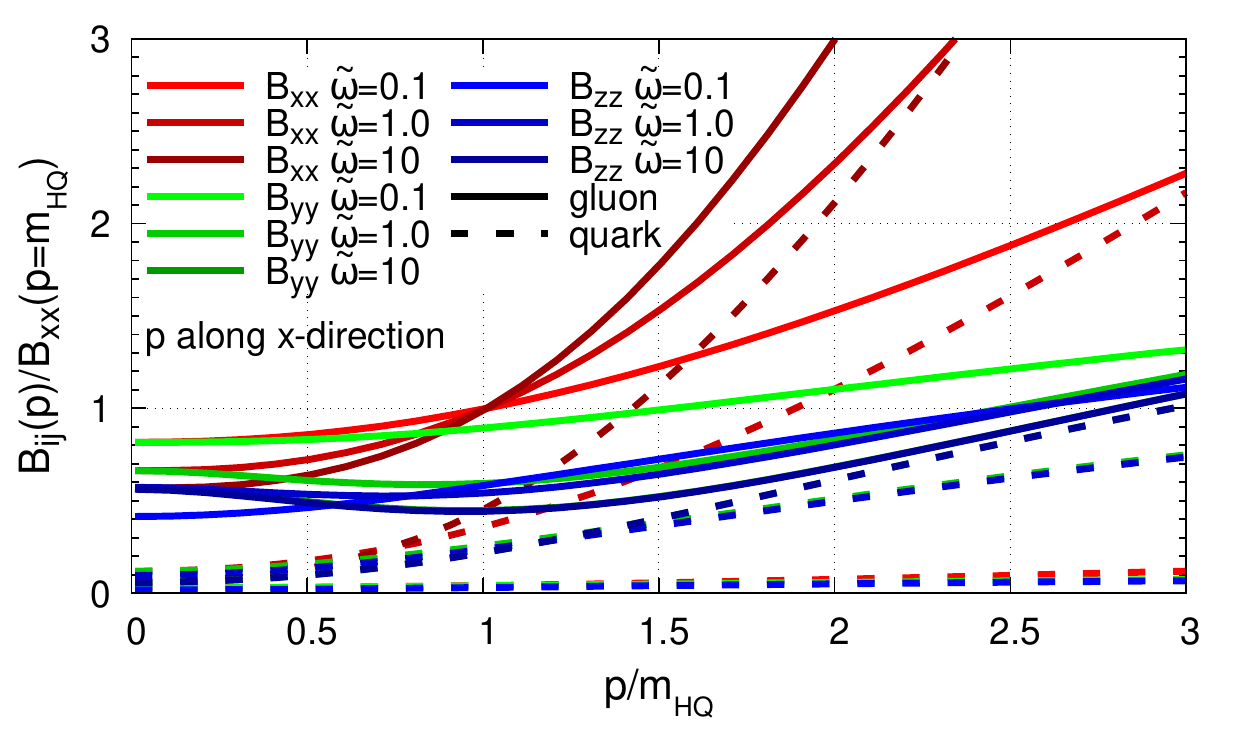} 
	\caption{Drag and diffusion coefficients $A_{x}(\vec{p})$, $B_{xx}(\vec{p})$ (red), $B_{yy}(\vec{p})$ (green), $B_{zz}(\vec{p})$ (blue) for gluon and quark (with factor 2$N_f$), as a function of rescaled momentum $p/m_{HQ}$. Coefficients are normalized by either $A_x(p=m_{\rm HQ})$ or $B_{xx}(p=m_{\rm HQ})$. The default coupling $\lambda=10$ is in use. The early-time coefficients are in lighter colors and the late-time coefficients are in darker colors.}
	\label{fig:AB_p}
\end{figure}

We plot the time evolution of coefficients $A_{x}$, $B_{xx}$, $B_{yy}$, $B_{zz}$ in Fig.~\ref{fig:AB_time} including their gluon and quark components.
The time evolution of the coefficients in terms of $\tilde{\omega}$ roughly features a pow-law behavior.
The lower panel of Fig.~\ref{fig:AB_time} shows the isotropization, that $B_{yy}$ is deviated from $B_{zz}$ at an early time and is approaching $B_{zz}$ at a late time.
The increasing trends in quark contribution show in both panels, that the drag and diffusion coefficients contributed by quarks become comparable to gluon at a late time.
However, due to the Bose-enhancement and Fermi-blocking factors from quantum statistics, the quark contribution in the coefficients is not as significant as it is in the energy density of the QCD plasma as we see in Fig.~\ref{fig:Plasma_time}, where $e_{q}\simeq 2e_{g}$ at the later time.

The momentum dependencies of the coefficients are shown in Fig.~\ref{fig:AB_p} for various times, where we have normalized everything by either $A_x(p=m_{\rm HQ})$ or $B_{xx}(p=m_{\rm HQ})$ so that these two coefficients are fixed at the point $(1,1)$.
One finds a linear dependence of $A_{x}(p)/A_{x}(m_{\rm HQ})$
for $p/m_{\rm HQ}\lesssim  2$.
Isotropization is shown in the lower panel that $B_{yy}$ gradually approaches $B_{zz}$ at a later time.
One also finds $B_{yy}\simeq B_{xx}$ for $p\ll m_{\rm HQ}$ for all time.
The increasing trend of the quark contribution is also presented.

\begin{figure}[t!]
	\centering
	\includegraphics[width=0.45\textwidth]
	{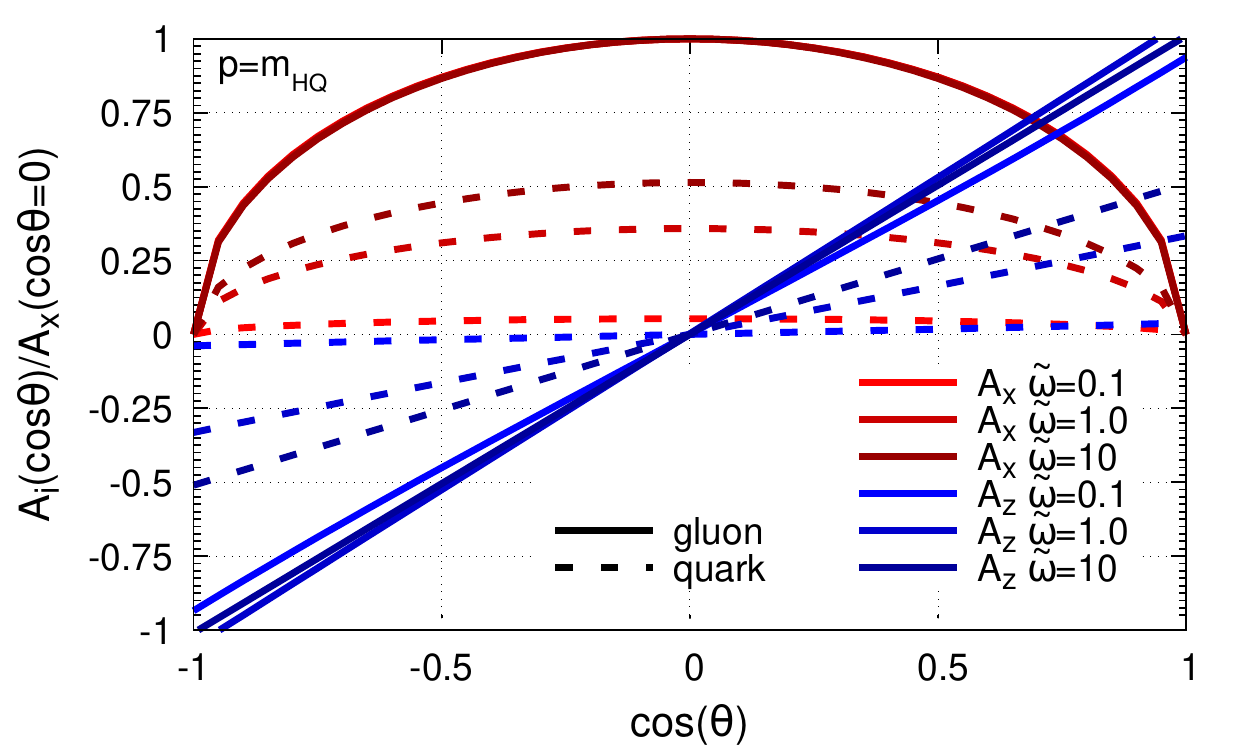} 
	\includegraphics[width=0.45\textwidth]
	{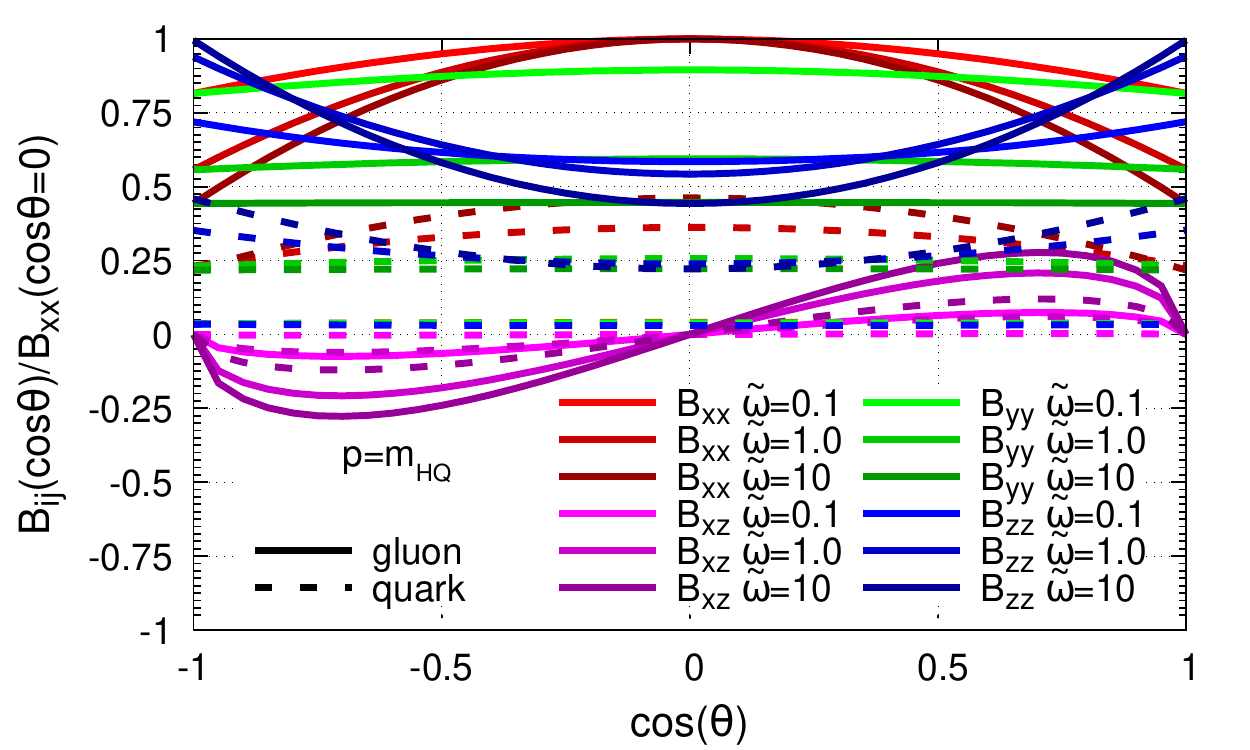} 
	\caption{Drag and diffusion coefficients $A_{i}(\vec{p},\tilde{\omega})$, $B_{ij}(\vec{p},\tilde{\omega})$ for gluon and quark (with factor 2$N_f$), as a function of angle $\cos(\theta)$ in x-z plane, normalized by values of $A_{x}$ and $B_{xx}$ when $\cos(\theta)=0$. The default coupling $\lambda=10$ is in use.}
	\label{fig:AB_theta}
\end{figure}

Now we release our constraints for the heavy quark momentum direction in the transverse plane.
Still, look at the typical momentum $p=m_{\rm HQ}$, but in the $x$-$z$ plane so that $\vec{p}=(p,\cos(\theta),\phi)=(p,\cos(\theta),0)$ in cylindrical coordinate. We therefore present the coefficients as a function of $\cos(\theta)=p_z/p$ in Fig.~\ref{fig:AB_theta}. 
Now the breaking symmetries in the integrals makes the vector $\vec{A}$ not necessarily along $\vec{p}$, resulting in many more nontrivial coefficients $A_{x,z}$, $B_{xx,xz,yy,zz}$, but they have vanishing points at certain angles.
For example, at $\cos(\theta)=0$, the heavy quark momentum is in the transverse plane and $A_{z}$ vanishes; while at $\cos(\theta)=\pm 1$, the heavy quark momentum is in the longitudinal plane, and $A_{x}$ vanishes.
The isotropization can be clearly seen, for example, the ratio of $A_z(\cos(\theta)=1)/A_x(\cos(\theta)=0)$ and $B_{zz}(\cos(\theta)=1)/B_{xx}(\cos(\theta)=0)$ only goes to 1 when the time is large and the medium becomes isotropic.
The off-diagonal diffusion coefficient $B_{xz}$ features a $\cos(\theta)$ dependence as $\braket{\braket{(p_x-p'_x)(p_x-p'_x)}}\sim \braket{\braket{\sqrt{1-\cos^2(\theta)}\cos(\theta)}}$ which vanishes at $\cos(\theta)=0,\pm 1$ for all time and peaks at $\cos(\theta)=1/\sqrt{2}$ when approaches the equilibrium.

\textbf{Rescaling of transport coefficients}

\begin{figure*}[t!]
	\centering
	\includegraphics[width=1.0\textwidth]
	{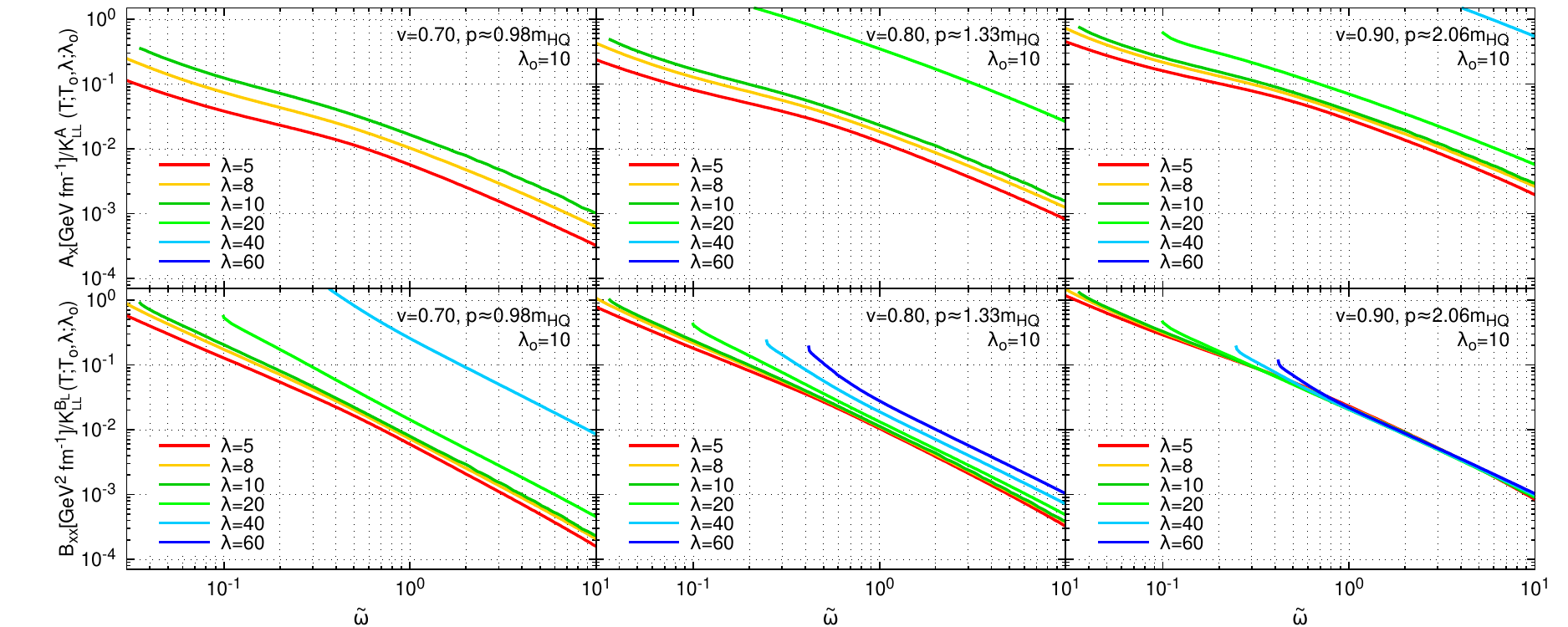} 
	\caption{Comparison of the time-dependent rescaled coefficients $A_{x}(T(\tilde{\omega}),\lambda)/\mathcal{K}_{\rm LL}^{A_{x}}(T(\tilde{\omega});T_o(\tilde{\omega}),\lambda;\lambda_o)$ and $B_{xx}(T(\tilde{\omega}),\lambda)/\mathcal{K}_{\rm LL}^{B_{xx}}(T(\tilde{\omega});T_o(\tilde{\omega}),\lambda;\lambda_o)$ at various couplings $\lambda=5,8,10,20,40,60$ with the LL $\mathcal{K}_{\square}$ factor parameterized as Eq.~(\ref{eq:K-LL-AB}) from~\cite{Moore:2004tg}. The values of the parameters for different velocities are taken from~\cite{Moore:2004tg} as well.}
	\label{fig:AB_Rescaling_LL}
\end{figure*}

With the kinetic theory simulated weakly coupled plasma and the universality of attractor valid even at strongly coupled regime, we may estimate the certain physical processes in the pre-hydrodynamic QCD plasma with different coupling strengths even at strong couplings.
Indeed, for any time convolution of a physical quantity $\mathcal{C}(\tau)$
\begin{eqnarray}
\label{eq:convolution}
\int_{\tau_0}^{\tau_{h}}\mathcal{C}(\tau)~d\tau_{\rm strong}\simeq \int_{\tau_0}^{\tau^*}\mathcal{C}(\tau)~d\tau_{\rm weak}\frac{\left(\eta/s\right)^{\frac{4}{3}}_{\rm strong}}{\left(\eta/s\right)^{\frac{4}{3}}_{\rm weak}},
\end{eqnarray}
where we have to fix the same initial time $\tau_0$ since the early time plasma is weakly coupled, until a hydrodynamization time $\tau_{h}$ in a strongly coupled plasma when $\tilde{\omega}\simeq 1-2$.
With attractor theory, one can estimate the hydrodynamization time in strongly coupled plasma $\tau_h$ from the time 
in weakly coupled plasma $\tau^*$ when $\tilde{\omega}\simeq 1-2$.
For the weakly coupled plasma $\lambda=10$ we have calculated, one has $(\eta/s)_{\rm weak}= 1$ and $\tau^{*}\simeq 100-275$$Q_s^{-1}$, which results in a hydrodynamization time $\tau_h\simeq 3.4-9.4 Q_s^{-1}\simeq 0.45-1.23$\,fm for strongly coupled plasma $(\eta/s)_{\rm strong}=1/4\pi$, or $\tau_h\simeq 8.6-23.7 Q_s^{-1}\simeq 1.13-3.12$\,fm for $(\eta/s)_{\rm strong}=1/2\pi$.

Strongly coupled plasmas featured by different coupling strengths also result in different values of the transport coefficients for heavy quarks. One may generically consider the rescaling of coefficients in plasma from the original coupling strength $\lambda_{o}$ to coupling strength $\lambda$, and from the original temperature $T_{o}$ to temperature $T$ as
\begin{eqnarray}
\mathcal{C}(\tau,\lambda)\simeq \mathcal{K}\left(T;T_o,\lambda;\lambda_o\right)\mathcal{C}(\tau,\lambda_o).
\label{eq:general_k}
\end{eqnarray}
Phenomenological studies of heavy flavor energy loss suggest a large overall rescaling factor $\mathcal{K}\simeq 5$ compared to the pQCD calculation~\cite{Rapp:2018qla} at $\alpha_s\simeq 0.3$ (comparable to $\lambda=4\pi\alpha_sN_c\simeq10$), empirically representing the non-perturbative effect.
This empirical value of the rescaling factor is however dominated by the contributions from the hydrodynamic period. The pre-hydrodynamic rescaling at higher temperatures would favor a smaller factor $\mathcal{K}\le 5$.
Although the rescaling for the conformal plasma with attractor theory can be safely rescaled to the strongly coupled regime, the rescalings for the drag and diffusion coefficients are theoretically non-trivial.
For instance, a rescaling from LO pQCD calculation would suggest an LL factor $\mathcal{K}_{\rm LL}\simeq \frac{\lambda^2}{\lambda_o^2}\frac{T^a}{T_o^a}\frac{b\ln(1/\lambda)+c}{b_o\ln(1/\lambda_o)+c_o}$, but it can only be safely restricted to the weakly-coupled regime.

More specifically, LL calculations for a heavy quark in a thermal background would suggest rescaling factors for the drag and diffusion coefficients as~\cite{Moore:2004tg}
\begin{eqnarray}
\nonumber
\mathcal{K}_{\rm LL}^{A}\simeq 
\frac{\lambda^2}{\lambda_o^2}\frac{T^2}{T_o^2}\left[
\frac{\ln(\frac{1}{\mu})+a_b+\frac{N_f}{2N_c}(\ln(\frac{1}{\mu})+a_f)}
{\ln(\frac{1}{\mu_o})+a_b+\frac{N_f}{2N_c}(\ln(\frac{1}{\mu_o})+a_f)}\right],\\
\nonumber
\mathcal{K}_{\rm LL}^{B_T}\simeq 
\frac{\lambda^2}{\lambda_o^2}\frac{T^3}{T_o^3}\left[
\frac{\ln(\frac{1}{\mu})+b_b+\frac{N_f}{2N_c}(\ln(\frac{1}{\mu})+b_f)}
{\ln(\frac{1}{\mu_o})+b_b+\frac{N_f}{2N_c}(\ln(\frac{1}{\mu_o})+b_f)}\right],\\
\mathcal{K}_{\rm LL}^{B_L}\simeq 
\frac{\lambda^2}{\lambda_o^2}\frac{T^3}{T_o^3}\left[
\frac{\ln(\frac{1}{\mu})+c_b+\frac{N_f}{2N_c}(\ln(\frac{1}{\mu})+c_f)}
{\ln(\frac{1}{\mu_o})+c_b+\frac{N_f}{2N_c}(\ln(\frac{1}{\mu_o})+c_f)}\right].
\label{eq:K-LL-AB}
\end{eqnarray}
with $\mu=\xi_g\frac{m_D}{T}=\xi_g\sqrt{\frac{\lambda(N_c+\frac{N_f}{2})}{3N_c}}=\xi_g\sqrt{\frac{\lambda}{2}}$ and $\xi_g=\frac{e^{5/6}}{2\sqrt{2}}$ an isotropic coefficient fitted to the HTL calculations.
The coefficients for the boson and the fermion sectors $a_b$, $a_f$, $b_b$, $b_f$, $c_b$, $c_f$ are velocity dependent.
The zero-velocity limit for a static heavy quark ($v=0$) gives the coefficients for diffusion $b_b=c_b=\ln2+\xi$ and $b_f=c_f=\ln4+\xi$ with $\xi=\frac{1}{2}-\gamma+\frac{\zeta'(2)}{\zeta(2)}\simeq -0.647$~\cite{Moore:2004tg}.
These LL factors break down at large couplings with a negative coefficient from the logarithmic term and deviate quite a bit from numerical results~\cite{Caron-Huot:2007rwy}. The appearance of the negativity when increasing the coupling strength is, however, delayed by a larger heavy quark velocity. As a consequence, a positive drag coefficient $A_i$ for $\lambda_o=10$ appears when $v\gtrsim 0.60$ from the LL factor, and a monotonously increasing $\mathcal{K}_{\rm LL}^{A}$ factor for $\lambda<\lambda_0=10$ occurs when $v\gtrsim 0.85$. Life is much easier for a weaker coupling $\lambda_o=5$, such that the positivity appears when $v\gtrsim 0.10$ and a monotonously increasing $\mathcal{K}_{\rm LL}^{A}$ factor requires $v\gtrsim 0.50$.

\begin{figure}[t!]
	\centering
	\includegraphics[width=0.45\textwidth]
	{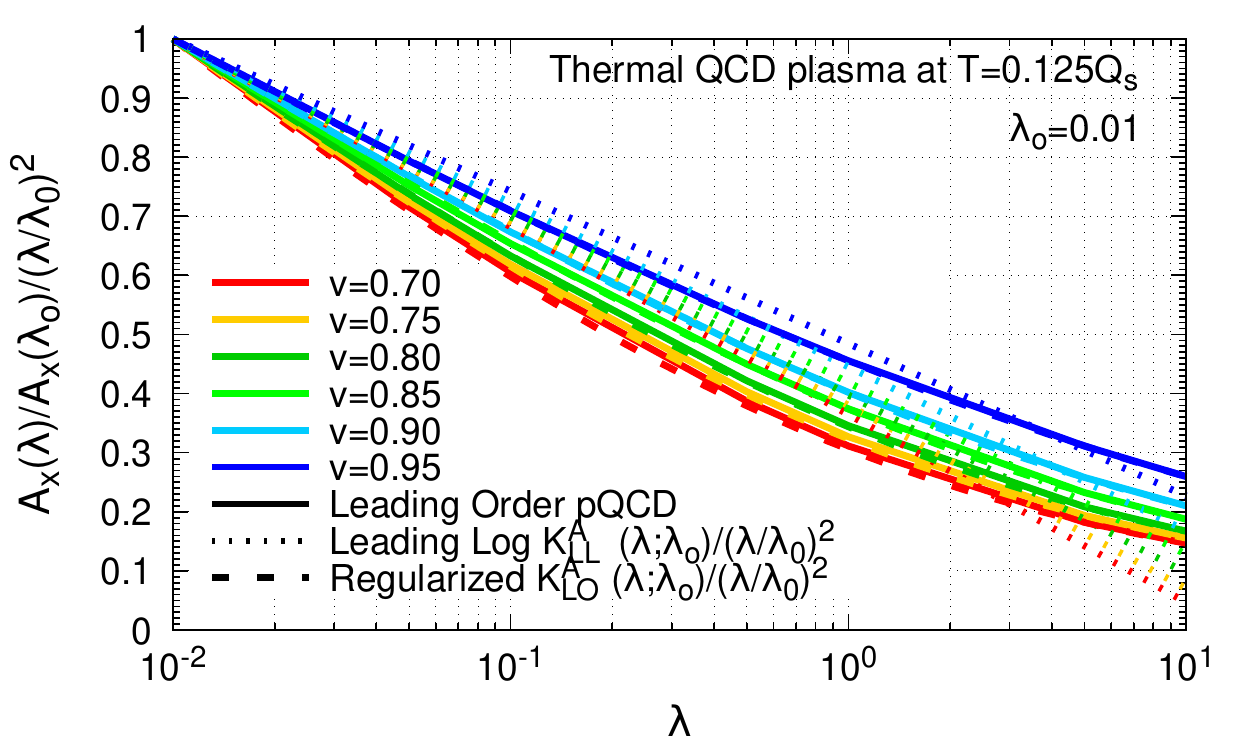} 
	\includegraphics[width=0.45\textwidth]
	{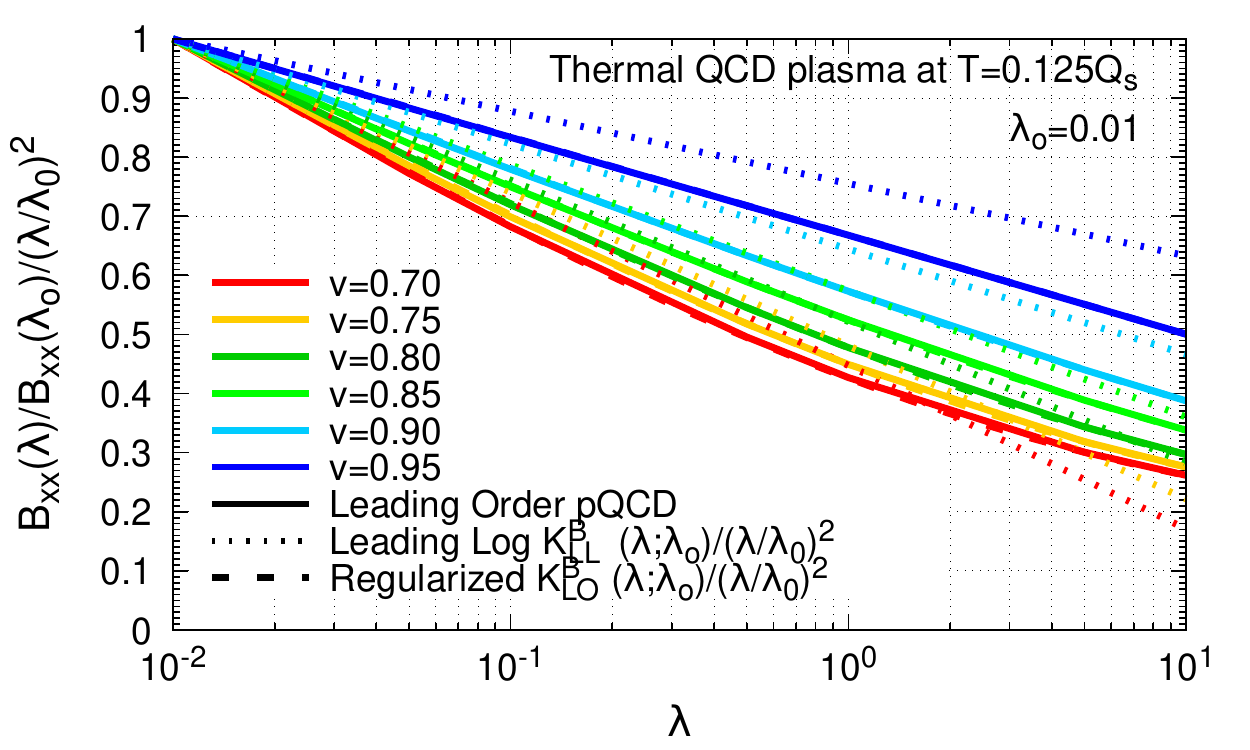} 
	\caption{Fitting the $\mathcal{K}_{\square}$ factors (dashed) in the square bracket term regularized as Eq.~(\ref{eq:K-LO-AB}) with numerical LO pQCD calculations (solid) in a thermal QCD plasma at fixed $T$ and various velocities. The fitting restricts itself at weak couplings ranging from $\lambda=0.01$ to $\lambda=10$. The LL $\mathcal{K}_{\square}$ factors without regulators are also presented (dotted).}
	\label{fig:AB_Rescaling_Fit}
\end{figure}

\begin{figure}[t!]
	\centering
	\includegraphics[width=0.45\textwidth]
	{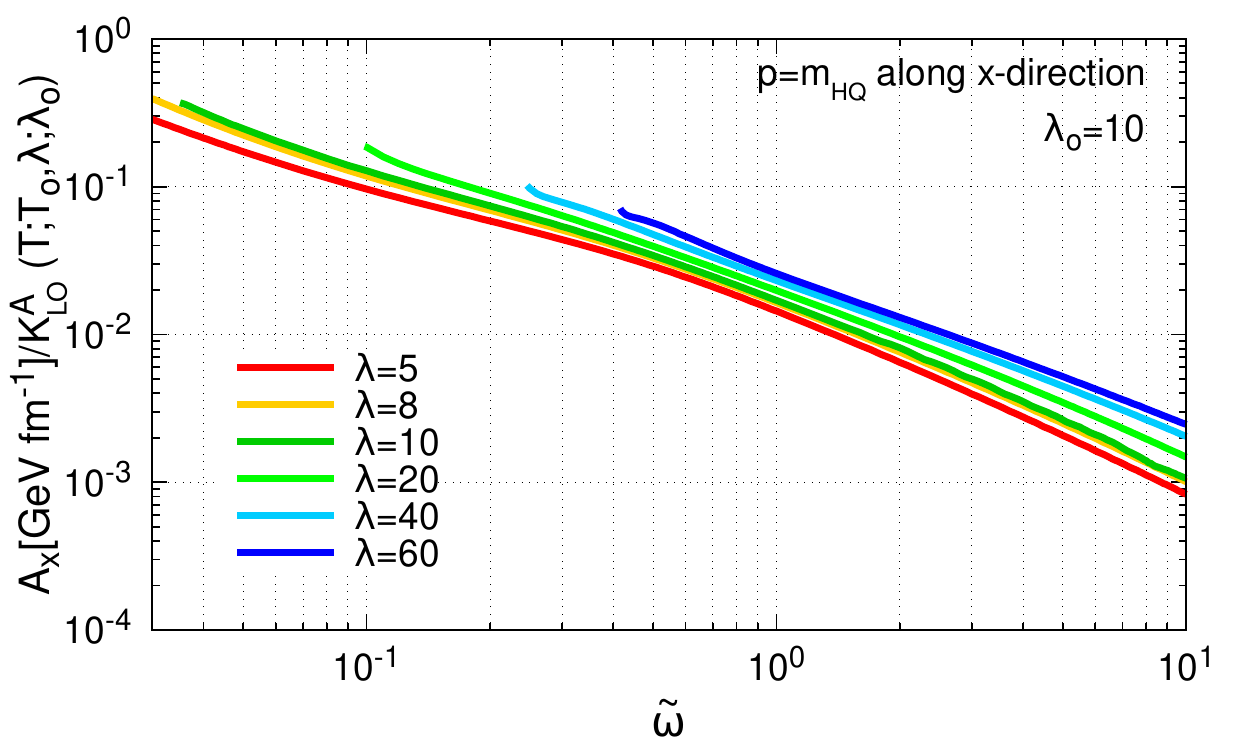} 
	\includegraphics[width=0.45\textwidth]
	{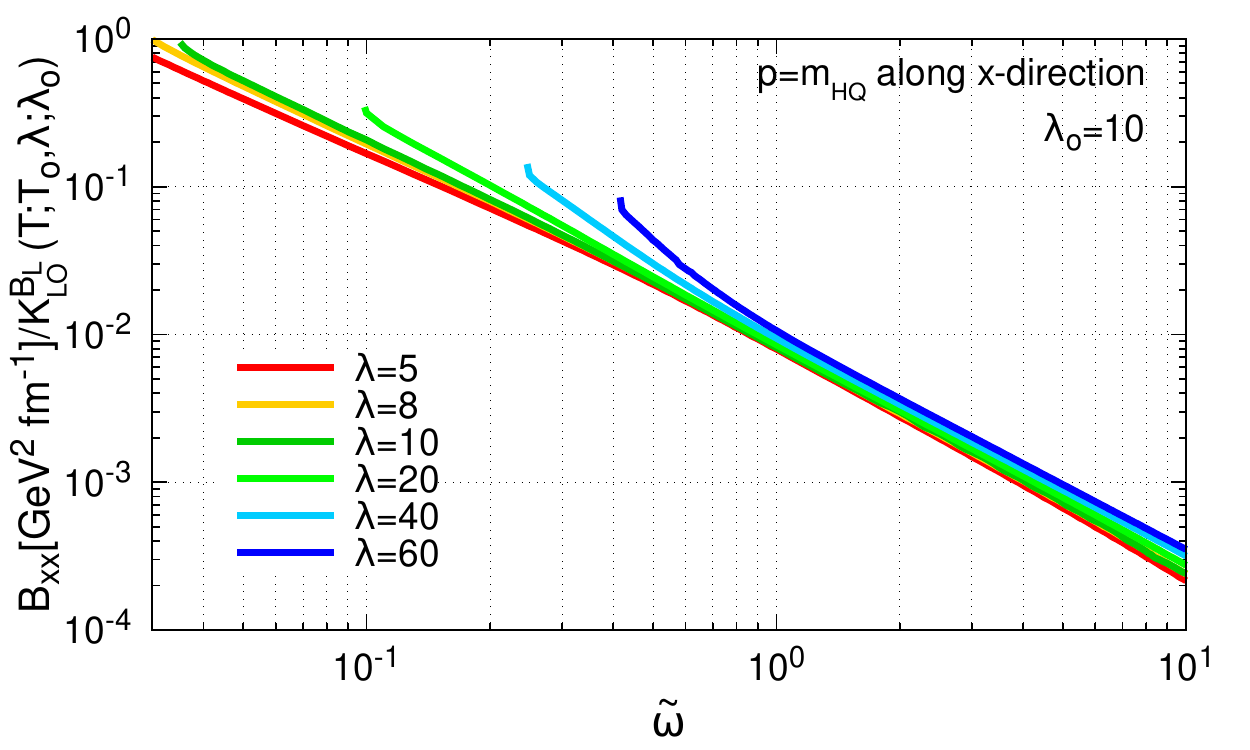} 
	\caption{Comparison of the time-dependent rescaled coefficients $A_{x}(T(\tilde{\omega}),\lambda)/\mathcal{K}_{\rm LO}^{A_{x}}(T(\tilde{\omega});T_o(\tilde{\omega}),\lambda;\lambda_o)$ and $B_{xx}(T(\tilde{\omega}),\lambda)/\mathcal{K}_{\rm LO}^{B_{xx}}(T(\tilde{\omega});T_o(\tilde{\omega}),\lambda;\lambda_o)$ at various couplings $\lambda=5,8,10,20,40,60$ with the regularized $\mathcal{K}_{\square}$ factor parameterized as Eq.~(\ref{eq:K-LO-AB}) from the LO pQCD.}
	\label{fig:AB_Rescaling}
\end{figure}

Due to the anisotropy in the pre-hydrodynamic plasma, we will only discuss $\mathcal{K}_{\rm LL}^{A_{x}}=\mathcal{K}_{\rm LL}^{A}\simeq\frac{\lambda^2}{\lambda_o^2}\frac{T^2}{T_o^2}\mathcal{K}_{\square}^{A}\simeq\frac{\lambda^2}{\lambda_o^2}\frac{(\eta/s)^{-2/3}}{(\eta/s)_o^{-2/3}}\mathcal{K}_{\square}^{A}$ and $\mathcal{K}_{\rm LL}^{B_{xx}}=\mathcal{K}_{\rm LL}^{B_{L}}\simeq\frac{\lambda^{2}}{\lambda_o^{2}}\frac{T^{3}}{T_o^{3}}\mathcal{K}_{\square}^{B_L}\simeq\frac{\lambda^{2}}{\lambda_o^{2}}\frac{(\eta/s)^{-1}}{(\eta/s)_o^{-1}}\mathcal{K}_{\square}^{B_L}$ with factor $\mathcal{K}_{\square}^{A}$ and $\mathcal{K}_{\square}^{B_L}$ the corresponding logarithmic terms in the square-bracket of Eq.~(\ref{eq:K-LL-AB}).
In Fig.~\ref{fig:AB_Rescaling_LL}, we present the time-dependent rescaled coefficients $A_{x}(T(\tilde{\omega}),\lambda)/\mathcal{K}_{\rm LL}^{A_{x}}(T(\tilde{\omega});T_o(\tilde{\omega}),\lambda;\lambda_o)$ and $B_{xx}(T(\tilde{\omega}),\lambda)/\mathcal{K}_{\rm LL}^{B_{xx}}(T(\tilde{\omega});T_o(\tilde{\omega}),\lambda;\lambda_o)$ 
calculated at various couplings, in comparison to our default coupling $\lambda_0=10$ (and $\mathcal{K}=1$ for $\lambda=10$) for $v=0.70,0.80,0.90$. 
It shows that the LL factors give close and presumably convergent rescaling results at weaker couplings, while the rescaling results diverge quickly at stronger couplings. 
A next-leading-order (NLO) correction cannot amend this due to poor convergence of the perturbative expansion~\cite{Caron-Huot:2007rwy} and it clearly presents the breakdown of perturbation theory at strong couplings.
It also appears that the LL factors rescale better at larger velocities. For example, at $v=0.90$, the diffusion coefficients have almost perfect rescalings.
It is conceivable that at the limit of $v\to 1$, the heavy quark becomes a high-energy jet and the perturbation theory is well valid. 
However, at such a large velocity, radiational processes dominate over collisional processes.
One may notice that some rescaled curves for large couplings are missing at low velocity due to negative values of the LL factors attributed by the $\ln(1/\mu)$ term.

The negativity from the logarithmic term of the LL factors in Eq.~(\ref{eq:K-LL-AB}) is due to an infrared cut of the parton momentum down to $m_D$, which is not a consistent treatment at strong couplings. 
If one does not impose the parton momentum to be much larger than the screening mass $k\simeq T\gg m_D\simeq gT$ or $g \ll 1$, one would expect an effective regulator in the logarithmic term.
Instead of the LL factors, we can employ a parameterization of the $\mathcal{K}$ factors with regulators $\tilde{a}_b$, $\tilde{a}_f$, $\tilde{b}_b$, $\tilde{b}_f$, $\tilde{c}_b$, $\tilde{c}_f$
\begin{eqnarray}
\nonumber
\mathcal{K}_{\rm LO}^{A}\simeq 
\frac{\lambda^2}{\lambda_o^2}\frac{T^2}{T_o^2}\left[
\frac{\ln(\frac{1}{\mu}+\tilde{a}_b)+a_b+\frac{N_f}{2N_c}(\ln(\frac{1}{\mu}+\tilde{a}_f)+a_f)}
{\ln(\frac{1}{\mu_o}+\tilde{a}_b)+a_b+\frac{N_f}{2N_c}(\ln(\frac{1}{\mu_o}+\tilde{a}_f)+a_f)}\right],\\
\nonumber
\mathcal{K}_{\rm LO}^{B_T}\simeq 
\frac{\lambda^2}{\lambda_o^2}\frac{T^3}{T_o^3}\left[
\frac{\ln(\frac{1}{\mu}+\tilde{b}_b)+b_b+\frac{N_f}{2N_c}(\ln(\frac{1}{\mu}+\tilde{b}_f)+b_f)}
{\ln(\frac{1}{\mu_o}+\tilde{b}_b)+b_b+\frac{N_f}{2N_c}(\ln(\frac{1}{\mu_o}+\tilde{b}_f)+b_f)}\right],\\
\nonumber
\mathcal{K}_{\rm LO}^{B_L}\simeq 
\frac{\lambda^2}{\lambda_o^2}\frac{T^3}{T_o^3}\left[
\frac{\ln(\frac{1}{\mu}+\tilde{c}_b)+c_b+\frac{N_f}{2N_c}(\ln(\frac{1}{\mu}+\tilde{c}_f)+c_f)}
{\ln(\frac{1}{\mu_o}+\tilde{c}_b)+c_b+\frac{N_f}{2N_c}(\ln(\frac{1}{\mu_o}+\tilde{c}_f)+c_f)}\right].\\
\label{eq:K-LO-AB}
\end{eqnarray}
Fitting the numerical LO pQCD calculations with these regularized factors
$\mathcal{K}_{\rm LO}\simeq\frac{\lambda^2}{\lambda_o^2}\frac{T^a}{T_o^a}\mathcal{K}_{\square}$ presented as Fig.~\ref{fig:AB_Rescaling_Fit}, we may have a better handle of the rescaling for larger couplings. The rescaling results with the regularized factor are presented for $p=m_{\rm HQ}$, $v\simeq 0.71$ in Fig.~\ref{fig:AB_Rescaling}. A nice rescaling is shown even at strong couplings.

Fast thermalization of the QGP requires the plasma to be strongly coupled when close to the hydrodynamic limit, while the perturbative calculations fail at large couplings. Although the regularized LL factors from the LO pQCD suggested by Eq.~(\ref{eq:K-LO-AB}) converge the large coupling rescalings, it is still an artificial treatment dropping higher-order corrections.
Rescaling of heavy quark drag and diffusion coefficients at large `tHooft couplings $\lambda=g^2N_c\gg 1$ might be achieved by the AdS/CFT correspondence, which suggests $B_{ij}\sim \sqrt{\lambda}T^{3}$~\cite{Herzog:2006gh,Casalderrey-Solana:2006fio}. Indeed, the non-polynomial rescaling from the AdS/CFT correspondence clearly indicates a non-perturbative effect. 

However, the smooth transition of the rescaling form from the weakly coupled pQCD polynomial results to the strongly coupled AdS/CFT non-polynomial results may not be possible to construct since their picture of heavy quark diffusion is very different. This aspect is out of the scope of the current study and we use the generic factors $\mathcal{K}^A\simeq\frac{\lambda^{\alpha}}{\lambda_0^{\alpha}}\frac{T^2}{T_0^2}\mathcal{K}_{\square}^{A}\simeq\frac{\lambda^{\alpha}}{\lambda_0^{\alpha}}\frac{(\eta/s)^{-2/3}}{(\eta/s)_0^{-2/3}}\mathcal{K}_{\square}^{A}$ 
and $\mathcal{K}^B\simeq\frac{\lambda^{\alpha}}{\lambda_0^{\alpha}}\frac{T^3}{T_0^3}\mathcal{K}_{\square}^{B}\simeq\frac{\lambda^{\alpha}}{\lambda_0^{\alpha}}\frac{(\eta/s)^{-1}}{(\eta/s)_0^{-1}}\mathcal{K}_{\square}^{B}$ with $\alpha\in \mathbb{R}^{+}$ representing some undetermined exponent in the following discussions.

\textbf{Phenomenological consequences}

Since we rescale from $\lambda_0=10$ which has $(\eta/s)_0=1$, we have accordingly, the energy loss and diffusion of heavy quarks at different couplings $\lambda$
\begin{eqnarray}
\label{eq:loss}
\left<\frac{\Delta p_i}{p_0}\right>_{\rm loss}&\simeq&\int_{\tau_0}^{\tau^{*}}-\frac{A_{i}(\vec{p},\tau)}{p_0}d\tau\left(\frac{\eta}{s}\right)^{\frac{4}{3}}\mathcal{K}^{A},\\
\label{eq:diff}
\left<\frac{\Delta p_i \Delta p_j}{p_0^2}\right>_{\rm diff}&\simeq&\int_{\tau_0}^{\tau^{*}}\frac{B_{ij}(\vec{p},\tau)}{p_0^2}\delta_{ij}d\tau\left(\frac{\eta}{s}\right)^{\frac{4}{3}}\mathcal{K}^{B}.
\end{eqnarray}
Focusing on the transverse plane and assuming the heavy quark initial momentum to be $p_0\lesssim 2 m_{HQ}$ in the x-direction, one has roughly a linear momentum dependence on the drag coefficient $A_x(\vec{p},\tau)\simeq \frac{A_x(m_{\rm HQ},\tau)}{m_{\rm HQ}}p_x$.
The infinitesimal form of Eq.~(\ref{eq:loss}) is 
\begin{eqnarray}
	\left<\frac{dp_x}{p_0}\right>_{\rm loss}
	&=&-\frac{A_{x}(\vec{p},\tau)}{p_0}d\tau\left(\frac{\eta}{s}\right)^{\frac{4}{3}}\mathcal{K}^{A}\\
	\nonumber
	&\simeq&-\frac{A_{x}(m_{\rm HQ},\tau)}{m_{\rm HQ}}\frac{p_x}{p_0}d\tau\left(\frac{\eta}{s}\right)^{\frac{4}{3}}\mathcal{K}^{A},
\end{eqnarray}
and the time convolution gives
\begin{eqnarray}
	&-&\int_{\tau_0}^{\tau^*}\frac{A_{x}(m_{\rm HQ},\tau)}{m_{\rm HQ}}d\tau\left(\frac{\eta}{s}\right)^{\frac{4}{3}}\mathcal{K}^{A}\\
	\nonumber
	&\simeq&
	\int_{p_0}^{p_0-\Delta p_x}\left<\frac{dp_x}{p_x}\right>_{\rm loss}=\ln\left(1-\left<\frac{\Delta p_x}{p_0}\right>_{\rm loss}\right).
\end{eqnarray}
Now we arrive at
\begin{eqnarray}
\label{eq:loss_final}
\left<\frac{\Delta p_x}{p_0}\right>_{\rm loss}&\simeq& 1-e^{-\int_{\tau_0}^{\tau^{*}}\frac{A_{x}(m_{\rm HQ},\tau)}{m_{\rm HQ}}d\tau\left(\frac{\eta}{s}\right)^{\frac{4}{3}}\mathcal{K}^{A}}\\
&\simeq& 1-e^{-\int_{\tau_0}^{\tau^{*}}\frac{A_{x}(m_{\rm HQ},\tau)}{m_{\rm HQ}}d\tau\left(\frac{\eta}{s}\right)^{\frac{2}{3}}\left(\frac{\lambda}{10}\right)^{\alpha}\mathcal{K}_{\square}^{A}}.
\end{eqnarray}
Numerical evaluations for $\tau^*$ give
\begin{eqnarray}
\left<\frac{\Delta p}{p_0}\right>_{\rm loss}\simeq 1-e^{-0.38(\frac{\eta}{s})^{\frac{2}{3}}\left(\frac{\lambda}{10}\right)^{\alpha}\mathcal{K}_{\square}^{A}}~{\rm up~to}~\tilde{\omega}=1\\
\left<\frac{\Delta p}{p_0}\right>_{\rm loss}\simeq1-e^{-0.56(\frac{\eta}{s})^{\frac{2}{3}}\left(\frac{\lambda}{10}\right)^{\alpha}\mathcal{K}_{\square}^{A}}~{\rm up~to}~\tilde{\omega}=2
\end{eqnarray}
before the plasma reaches the hydrodynamic stage.
With a proper rescaling factor $\mathcal{K}_{\square}$ presumably from some non-perturbative numerical methods, the above formula is useful to simply estimate the heavy quark energy loss in the pre-hydrodynamic stage in HICs.

\textbf{Conclusions \& Outlook}

In this article, we calculate the heavy quark drag and diffusion coefficients in a weakly coupled pre-hydrodynamic QCD plasma from a first principle and the state-of-the-art QCD EKT solver. 
We present the time, momentum, and angular dependencies of the coefficients $A_{i}$, $B_{ij}$ with all indices.
With arguments from the attractor theory, we provide a simple formula to evaluate the heavy quark energy loss in pre-hydrodynamic plasma with different coupling strengths. 
As a first step towards this goal, we study the rescaling of transport coefficients with the LL factors. Although a trend of convergence shows at weak couplings, the rescalings augmented with LL factors diverge at strong couplings due to the failure of perturbation theory. Releasing the restriction of the infrared cut for parton momentum down to the screening mass, one may employ a regulator in the logarithmic terms of LL factors and fit the regulator to the LO pQCD calculations, which avoids the negativity from native LL factors and leads to convergent rescalings for large couplings.

One needs to keep in mind that the attractor rescaling of the QCD plasma is valid at both weak and strong couplings. The QCD EKT simulations of the time and temperature profiles of the plasma appear to satisfy this rescaling even up to a large coupling $\lambda=60$ whose corresponding $\eta/s\simeq 0.084$ is already close to the holographic lower bound, regardless of the failure of the semiclassical kinetic theory at strong couplings.
In a realistic pre-hydrodynamic stage in HICs, the rapid drop in temperature and fast thermalization of the plasma favors a quick transition from weakly coupled to strongly coupled, which may not be easy to simulate by kinetic theory or other theories. Attractor theory provides a simple way without any complex simulation to perform a rescaling in the weakly/strongly coupled transition for the QCD plasma.

However, the difficulty in calculating the heavy quark transport coefficients may come from the invalidity of different theories at different scales and coupling strengths. 
A generic rescaling from a weakly coupled regime to a strongly coupled regime is theoretically non-trivial. Still, it is conceivable upon incorporating non-perturbative approaches like the T-matrix calculation~\cite{vanHees:2007me} or functional renormalization group (FRG) for calculating the transport coefficients, which we leave for future studies.  

\begin{acknowledgments}
\textbf{Acknowledgement}

The author thanks Kirill Boguslavski, Florian Lindenbauer, Meijian Li, Sören Schlichting, Haitao Shu, Bin Wu, and anomalous reviewers for helpful discussions. 
The author is supported by Xunta de Galicia (Centro singular de investigacion de Galicia accreditation 2019-2022), European Union ERDF, the “Maria de Maeztu” Units of Excellence program under project CEX2020-001035-M, the Spanish Research State Agency under project PID2020-119632GB-I00, and European Research Council under project ERC-2018-ADG-835105 YoctoLHC. 
The author also acknowledges the computational resources supported by LUMI-C supercomputer, under The European High Performance Computing Joint Undertaking grant EHPC-REG-2022R03-192 Non-equilibrium Quark-Gluon Plasma.
\end{acknowledgments}

\bibliography{ref}
\end{document}